\documentclass[sn-mathphys-num]{sn-jnl}

\usepackage{amsmath,amssymb,amsfonts}%
\usepackage{amsthm}%
\usepackage{mathrsfs}%
\usepackage[title]{appendix}%
\usepackage{xcolor}%
\usepackage{textcomp}%
\usepackage{manyfoot}%
\usepackage{booktabs}%
\usepackage{algorithm}%
\usepackage{algorithmicx}%
\usepackage{algpseudocode}%
\usepackage{listings}%
\usepackage{subcaption}%
\usepackage{multirow}
\usepackage{rotating}
\usepackage{tabularray}
\usepackage{array}
\usepackage{tabularx}
\usepackage{hyperref}

\theoremstyle{thmstyleone}%
\theoremstyle{thmstyletwo}%

\theoremstyle{thmstylethree}%

\begin{document}

\title[Article Title]{TPFL: Tsetlin-Personalized Federated Learning with Confidence-Based Clustering}
\author*[1]{\fnm{Rasoul Jafari} \sur{Gohari}}\email{rjafari@math.uk.ac.ir}
\author[2, 3]{\fnm{Laya} \sur{Aliahmadipour}}\email{l.aliahmadipour@uk.ac.ir}
\author[4]{\fnm{Ezat} \sur{Valipour}}\email{valipour@uk.ac.ir}
\affil*[1]{\orgdiv{Department of Computer Science}, \orgname{Shahid Bahonar University of Kerman}, \orgaddress{\city{Kerman}, \country{Iran}}}
\affil[2]{\orgdiv{Department of Computer Science}, \orgname{Faculty of Mathematics and Computer, Shahid Bahonar University of Kerman}, \orgaddress{\city{Kerman}, \country{Iran}}}
\affil[3]{\orgdiv{Department of Computer Science}, \orgname{Faculty of Mathematical Sciences and Statistics, Malayer University}, \orgaddress{\city{Malayer}, \country{Iran}}}
\affil[4]{\orgdiv{Department of Applied Mathematics}, \orgname{Faculty of Mathematics and Computer, Shahid Bahonar University of Kerman}, \orgaddress{\city{Kerman}, \country{Iran}}}
\abstract{The world of Machine Learning (ML) has witnessed rapid changes in terms of new models and ways to process users' data. The majority of work that has been done is focused on Deep Learning (DL) based approaches. However, with the emergence of new algorithms such as the Tsetlin Machine (TM) algorithm, there is growing interest in exploring alternative approaches that may offer unique advantages in certain domains or applications. One of these domains is Federated Learning (FL), in which users' privacy is of utmost importance. Due to its novelty, FL has seen a surge in the incorporation of personalization techniques to enhance model accuracy while maintaining user privacy under personalized conditions. In this work, we propose a novel approach called TPFL: Tsetlin-Personalized Federated Learning, in which models are grouped into clusters based on their confidence towards a specific class. In this way, clustering can benefit from two key advantages. Firstly, clients share only what they are confident about, resulting in the elimination of wrongful weight aggregation among clients whose data for a specific class may not have been enough during the training. This phenomenon is prevalent when the data are non-Independent and Identically Distributed (non-IID). Secondly, by sharing only weights towards a specific class, communication cost is substantially reduced, making TPLF efficient in terms of both accuracy and communication cost. The TPFL results were compared with 5 other baseline methods; namely FedAvg, FedProx, FLIS, IFCA and FedTM. The results demonstrated that TPFL performance was better than baseline methods with 98.94\% accuracy on MNIST, 98.52\% accuracy on FashionMNIST and 91.16\% accuracy on FEMNIST dataset.}




\maketitle

\section{Introduction}\label{sec1}

Today's world is heavily driven by numerous Machine Learning (ML) algorithms and it is almost impossible to ignore its pervasive existence in our digital lives. This widespread adoption of ML applications equals huge amounts of never-before-seen data from various types of sources, whether it is a face detection camera or a recommendation based on a query in a search engine {w1}. This phenomenon in itself raises critical concerns in regards to user privacy as well as data security \cite{w2}. Traditional ML techniques rely on centralized approaches, where users data is displaced and moved to where the algorithm resides \cite{w3}. The technique suffers from numerous disadvantages, including arduous maintenance of large amounts of data which is produced in mass from various parties, communication cost of data transfer from parties to the centralized server, and finally the most important one, serious privacy risks againsts users data \cite{w4}. 

One of the recent promising solutions for tackling such hurdles is Federated Learning (FL). This new paradigm has enabled many to collaboratively train models on decentralized devices while safeguarding users privacy \cite{w5}. This novel technique prevents users' data from being displaced and instead moves the algorithm where data resides \cite{w6}. FL paves the way for the creation of on-device models that are trained on local data. As a result, instead of sharing the users' data, models' parameters are shared with a centralized aggregator where models are combined to create a single, high-performance global model. This new strategy enables each local model to achieve accurate results by focusing its learning process on user-specific data distribution. In other words, the local model is tailored to local data \cite{w7}. In this way, maintenance of the users data in a centralized server is eliminated and privacy risks no longer pose a serious threat. However, this paradigm has not been perfect in recent years. One of the major disadvantages of FL algorithms is struggling to reach a high degree of accuracy in the aggregator. This struggle in the aggregator becomes evident when the data is non-Independent and Identically Distributed (non-IID)\cite{w8}.

Non-IID data is undoubtedly an important challenge in the FL settings, which usually ends up to be the global model's biggest reason for low accuracy in comparison to local models. This phenomenon stems from heterogeneous data distribution among participating clients. Combining models in the aggregator in the face of data heterogeneity is the main problem that prevents global model from achieving high-performance results. For this reason, Personalized Federated Learning (PFL) has been a fruitful strategy to tackle this issue \cite{w9}. PFL takes data heterogeneity into consideration and eliminates the one-size-fits-all assumption so that the aggregation of the personalized model of each participant can result in a global model that reflects the data distribution of each device. In other words, a model that works well on one's device will most probably have unacceptable performance on another client's device since their data distribution is completely different from one another. Therefore, if all models are simply aggregated without any considerations, accuracy is significantly reduced in the global model \cite{w10}. 

One of PFL strategies to keep the personalized preferences of clients intact during the aggregation is multi-center FL, in which similar clients are partitioned into clusters. Thus, unlike single-center FL where there is only one single global model for all the clients, with multi-center FL we can take personalization into account and group similar clients into their corresponding cluster. This multi-center partitioning establishes numerous clusters, each responsible for weight aggregation of its corresponding clients. One of the best ML algorithms that can be used in FL settings is the Tsetlin Machine (TM). The TM algorithm is a great choice compared to Deep Learning (DL) when it comes to interpretability and explainability. This phenomenon stems from the fact that the TM relies on a voting mechanism that enables the TM models to be more transparent in comparison with DL models that possess a black-box nature \cite{w11}. In this paper, we present a novel multi-center FL method that, to the best of our knowledge, leverages the Tsetlin Machine (TM) algorithm for the first time in the context of PFL, which we call Tsetlin-Personalized Federated Learning (TPFL). This innovative approach is designed to perform effectively with both IID and non-IID data in a few-shot learning fashion. In essence, the clustering mechanism of TPFL is directly based on each client's confidence towards a specific class of data. In fact, clients generally in an FL setting tend to be trained on data that they generate locally. Therefore, a client that is trained on images of odd numbers tends to have more confidence towards those classes of data. And similarly a client with images of even numbers is confident about what it had been trained on; a range of even numbers. As a result, the more data the local model is trained on, the more confident it becomes towards that particular class. Our novelty specially holds true when FL clients data distribution is non-IID. In other words, when some random clients are only trained on data whose distribution is heavily skewed towards a similar class of data, they must be confident and consequently partitioned into the same cluster.

A summary of our contributions are as follows:

\begin{itemize}
  \item We introduce a novel personalization approach based on multi-center clustering using the TM algorithm that is we call TPFL.
  \item We provide a client-side algorithm that shows how a client participates in each round of training by sharing only a single weight vector per round. This vector corresponds to the class towards which the client's confidence is highest.
  \item We also provide an algorithm in the aggregator side for the aggregation of clients weights. This indicates that the most number of clusters that we may have is equal to the number of classes.
  \item We evaluate the TPFL method under 5 different experimental setups on 3 datasets, showing the effectiveness of TPFL.
  \item We compare the TPFL result with 6 other baseline methods. This comparison covers not only the methods accuracy but also their communication costs as well.
\end{itemize}

The organization of this paper is as follows: Section \ref{Section2} provides a review on related works that proposed a novel method for PFL, Section \ref{Section3} discusses the motivation behind this work, Section \ref{Section4} goes over the preliminaries, Section \ref{Section5} provides a comprehensive explanation on the proposed TPFL method and Section \ref{Section6} provides the obtained results of our implementation of TPFL and other baseline methods.

\section{Related Works}\label{Section2}

Since the advent of FL-based architecture, different approaches have opted for PFL to increase the efficiency and accuracy of FL models. Ghosh et al. in \cite{w12} introduce a one-shot clustering algorithm called Iterative Federated Clustering Algorithm (IFCA) to address the challenges in FL clustering using Empirical Risk Minimizers (ERM) from FL clients. They also discuss practical aspects of implementing IFCA, including a weight-sharing scheme that reduces communication costs between the aggregator and FL clients. The IFCA uses FEMNIST and MNIST for its work, which yields satisfactory accuracy in all experimental setups. Work of Briggs et al. introduces a novel method for training specialized ML models over distributed datasets enhanced by hierarchical clustering \cite{w13}. Their paper continues to add a clustering step within the federated learning protocol, which was clustering clients based on the similarity of their model weight updates. Another positive aspect of this work was comprehensive analysis of how hierarchical clustering impacts test set accuracy across different IID and non-IID data settings. Another valuable work for clustering-based PFL was the work of Ruan and Joe-Wong in \cite{w14}, which proposed FedSoft. Their contribution was based on proximal local updates. Unlike traditional hard clustering, where each client belongs to a single cluster, FedSoft allows clients to be associated with multiple clusters with varying degrees of membership. A proximal term is added to the local objective function of each client, ensuring the local models stay close to the global model, which as a result helps in mitigating the divergence issues often encountered in federated learning due to heterogeneous data distributions across clients.\\
The work of Long et al. brings forward an approach called Federated Stochastic Expectation Maximization (FeSEM) that solves the challenge of non-IID data using a framework with multi-center clusters for clients \cite{w15}. This approach partitions clients based on their data similarity using K-Means algorithm that is guided by the FeSEM algorithm. Effectively, each user is aligned with its corresponding global model and as a result, the loss is minimized in each cluster. The authors used Federated Extended MNIST (FEMNIST), FedCelebA datasets for their work and case study comparison to illustrate the effectiveness of their results. The authors however, did not show the communication cost of their model since one of the important criteria within FL settings is the communication cost between the clients and the aggregator, whether the FL setting is based on a single-center or a multi-center architecture. Work of Mansour et al. \cite{w16} proposed three distinct approaches to achieve better personalization in FL settings which are user clustering, data interpolation and model interpolation, respectively. Although the approaches are not combined, their comparison sheds a light on different aspects of personalization. The clustering scheme allows the FL architecture to be more scalable than the predecessor approaches. Their benchmark datasets were EMNIST-62 as well as synthetic data that was used for the initial evaluation and understanding of the algorithms' behavior.  Sattler et al. proposed a novel model-agnostic clustering algorithm for clients whose distributions are similar when it comes to non-IID data \cite{w17}. Their unique approach is the adaptiveness of their framework, which adjusts the clusters as more data becomes available. They used MNIST, CIFAR-10 and Fashion-MNIST (FMNIST) datasets for their framework. Gong et al. proposed another framework for clustering clients based on non-IID data \cite{w18}. Their novel strategies include a local training adjustment strategy as well as an adaptive weighted strategy for client clustering based on voting. Model accuracy was improved and communication overhead was reduced by evaluating the model on MNIST, FMNIST and CIFAR-10 datasets. \\
Morafah et al. have proposed yet another federated framework for non-IID data called Federated Learning by Inference Similarity (FLIS) \cite{w19}. Their clustering approach is based on inference similarity. Their approach does not force the number of clusters to be known prior to the cluster creation process. Their model-agnostic approach minimizes the need for frequent communication between clients and the central server. Their work focuses on MNIST, FMNIST and CIFAR-10. Cho et al. have proposed a framework called COMET that takes the local models' data as well as their architecture into consideration, which is likely to be a scenario for industrial FL architecture \cite{w20}. Moreover, their approach relied on knowledge distillation mechanism that allows clients to share only the logits (predictions) on a common unlabeled dataset. In addition to reduced communication overhead, their algorithm's improved accuracy outperforms some of the contenders in FL settings. Their work was trained on CIFAR-10 and CIFAR-100 datasets.
Work of Fraboni et al. aims to enhance client representativity and reduce the variance in the stochastic aggregation weights of clients. This framework is based on 1) sample size-based clustering, which groups clients based on the size of their datasets and 2) model similarity-based clustering, which groups clients based on the similarity of their model updates. The framework is trained on MNIST and CIFAR-10 datasets \cite{w21}. Work of Camara et al. proposes a clustered FL architecture specifically designed for unsupervised anomaly detection in large, heterogeneous networks of IoT devices. This approach addresses the challenges of data isolation, network overhead, and privacy concerns. Their results demonstrate the improvements in anomaly detection performance in addition to reduction of network overhead and improved global model convergence. Their work was evaluated on a benign IoT network traffic testbed called Gotham \cite{w22}. Work of Yin et al. introduced an efficient one-off clustered federated learning framework called FedEOC, which aims to improve the balance between accuracy and efficiency in federated learning by clustering clients only once and applying the clusters over multiple rounds. FedEOC leverages the learning-to-learn characteristic of meta-learning in order to quickly adapt to new tasks, ensuring that the initial model parameters are well-suited for all clients. The FedEOC framework was evaluated on the ml-100K and ml-1M dataset \cite{w23}. Li et al. put forward a new algorithm called Federated Learning with Soft Clustering (FLSC). This algorithm combines the strengths of soft clustering and the Iterative Federated Clustering Algorithm (IFCA). Contrary to IFCA, which uses non overlapping clusters, FLSC allows for overlapping clusters. This means that each client can contribute to multiple clusters, improving the utilization of local information \cite{w24}. Their work used MNIST and FMNIST datasets. He et al. proposed a framework that brings forward a novel layer-wise perturbation function that reduces the Local Differential Privacy (LDP) noise in comparison to relevant methods. As a result, communication overhead was decreased in Internet of Things (IoT) devices. MNIST, fashion-MNIST (FMNIST) and Federated Extended MNIST (FEMNIST) were used for the evaluation of their framework \cite{w25}. Finally, the work of Ghosh et al. addresses the challenges of robustness in a heterogeneous environment with Byzantine nodes. Their proposed method consists of a three-stage modular algorithm; 1) Worker nodes compute their Local Empirical Risk minimizers (ERMs) and send them to the center node, 2) The center node clusters these local ERMs to form distinct clusters and 3) Each cluster undergoes a Byzantine-tolerant distributed optimization process to refine the model parameters. Results with 3 different clustering strategies were presented, namely K-means (KM), Trimmed K-means (TKM), and K-geomedians (KGM). The results demonstrate that the proposed algorithm outperforms standard non-robust algorithms in both synthetic and real-world datasets. The real-world dataset that was used was Yahoo! Learning to Rank Dataset \cite{w26}.

\section{Motivation}\label{Section3}
The advent of Federated Learning (FL) has paved the way to the collaboration of training among multiple Artificial Intelligence (AI) models while maintaining user data privacy. Recent advancements within FL have led many to propose personalization techniques in order to increase the efficiency of FL models. Nevertheless, the potential use of other state-of-the-art algorithms like the TM algorithm is normally overlooked by the widespread use of DL models within the FL context. 

The main principle of the TM algorithm is its propositional logic and its voting mechanism. This inherent voting scheme provides a unique capability for calculating the confidence score--the higher the vote margin, the more confident the TM is in its prediction for a certain class. Not only the TM itself, but also its variants provide such an interpretable architecture. Recently, numerous other advancements have been proposed, such as the Convolutional Tsetlin Machine (CTM) that is specifically designed for image recognition tasks \cite{w27} and Regression Tsetlin Machine (RTM) that addresses continuous prediction problems \cite{w28}. This versatility illustrates the potential that the TM algorithm holds for edge and IoT devices within an FL setting. This becomes more interesting when computational resources are not abundant, paving the way for a more efficient and scalable approach to AI development. This is due to the fact that the TM's advantage compared to the DL models is its hardware-friendliness. As a result, the TM can be considered a prime candidate as a computationally-efficient algorithm.

Given the recent advancements in the TM, we are motivated to explore its potential within the context of PFL. We are particularly interested in comparing the communication efficiency as well as accuracy of the TM with other leading proposed frameworks. Recent research in FL setting suggests that the TM performs significantly well in pattern recognition tasks without requiring the substantial hardware resources needed by DL models \cite{w29}.

\section{Preliminaries}\label{Section4}

In this section, we aim to shed light on pivotal topics that are essential for the comprehension of our proposed TPFL method. We try to lay the foundation for a deeper understanding of our research approach, which is built upon two sections; namely,  TM algorithm and problem statement. 

\subsection{Tsetlin Machine}
\begin{figure}[h]\label{TM}
\centering
\includegraphics[width=0.7\textwidth]{"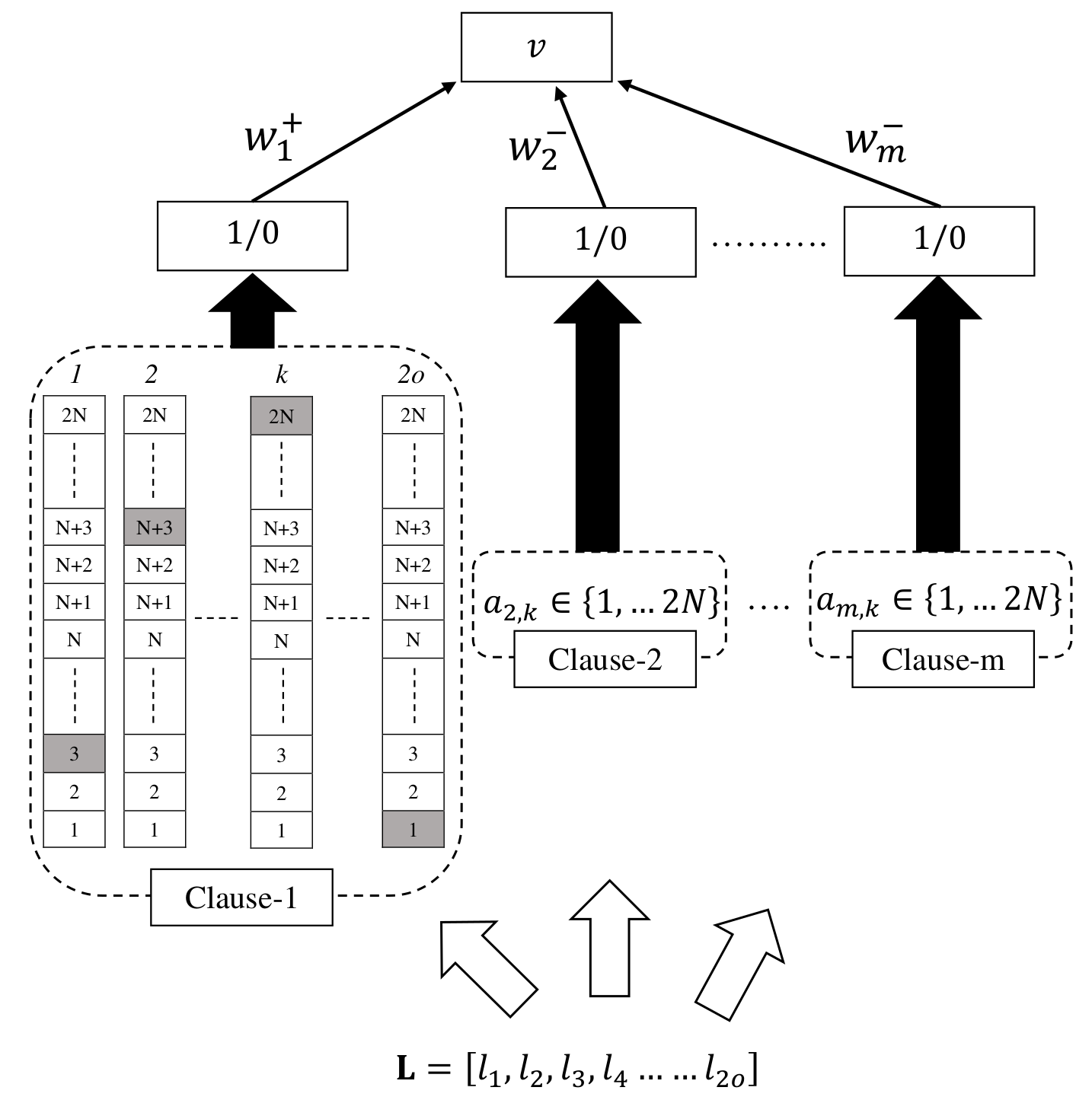"}
\caption{TM architecture \cite{w30}}\label{fig1}
\end{figure}

The TM algorithm is a relatively new ML algorithm that was introduced by Granmo in \cite{w30}. The basic principle components of the TM are based on binary, rule-based methods. This algorithm is composed of  Tsetlin Automatons (TA) whose foundation is based on states, action and reward-penalty feedback. The TM uses a collection of two-action TAs where a TA with $2N$ states decides its next action based on its current state. States from 1 to $N$ trigger Action 1, and states from $N+1$ to $2N$ trigger Action 2.The TAs direct interactions with an environment alters the states of the TAs based on a reward or penalty feedback. Reward feedback reinforces the action, while penalty feedback weakens it. Therefore, the TM completely differs from continuous-valued parameters and gradient-based optimization methods such as DL methods. However, similar to DL models, the TM algorithm is composed of multiple layers in order to learn from data. Figure \ref{fig1} demonstrates these layers. 

The TM starts with a literal vector $L = [x_1, x_2, \ldots, x_o, \neg x_1, \neg x_2, \ldots, \neg x_o]$, which includes all input features and their negations. The goal is to classify input feature vectors $X$ using $o$ propositional variables $x_k \in \{0, 1\}$ and their negations $\neg x_k$. This pattern allows the TM to carry out pattern recognition more efficiently since the absence of a feature can generally assist with the reward or penalty feedback.

Next layer is clause construction, in which a TM is comprised of $m$ conjunctive clauses. All clauses during the training receive the same literal vector $L$. Additionally, each clause holds a sub-pattern within itself that is associated with a class. Also, each clause includes the distinct literals indexed by $I_j \subseteq \{1, \ldots, 2o\}$. Each clause is equipped with two TAs per literal, which determine whether the literal should be included or excluded. States from 1 to $N$ decide exclusion, and states from $N+1$ to $2N$ decide inclusion. 

The even-indexed clauses of the TM are given a positive polarity $(c_j^+)$ while the odd-indexed clauses are given a negative polarity $(c_j^-)$. This division allows the even-indexed clauses to vote in favor of sub-patterns of output $y = 1$ and odd-indexed clauses to vote in favor of sub-patterns of output $y = 0$ for a binary classification task. 

Finally, the TM moves forward with the classification task by utilizing a weighted voting mechanism. The TM algorithm incorporates a weighting scheme to each and every clause using the following notation:

\begin{equation}
\hat{y} = 
\begin{cases} 
1 & \text{if } \sum_{j=1}^{n/2} W_j^{i, +}C_j^{i, +} - \sum_{j=1}^{n/2} W_j^{i, -}C_j^{i, -} \\ 
0 & \text{otherwise}.
\end{cases}	\label{eq1}
\end{equation}

in which $c_j^+$ is the vote summation of all positive-polarity clauses for class $i$ and $c_j^-$ is the vote summation of all negative-polarity clauses for class $i$. Also, $w_j^+$ is the weight of the $j^{th}$ clause with positive polarity for class $i$ and $w_j^-$ is the weight of the $j^{th}$ clause with negative polarity for class $i$ \cite{w31}.

\subsection{Problem Statement}
One of the main challenges in the FL settings is the heterogeneity of data among clients. We aim to enhance the accuracy of clients by introducing a novelty that allows clients to avoid sharing their knowledge if they had insufficient training data on that class. This avoidance allows client's weights to be aggregated only if they are confident about what they have learned. Therefore, clients in TPFL are grouped into clusters based on their confidence scores towards a specific class of the data.

Given a FL setting with $N$ clients and a dataset with $C$ classes, our goal is to train clients and aggregate their weights based on confidence towards an identical class. For each client $i$ where $i \in \{1, 2,  . . . , N\}$, we denote their confidence score towards class  $l$ as $conf_{i, l}$, where $l \in \{1, 2, . . . , C\}$.

Each client \( i \) is assigned to a cluster \( k \) where \( k \in \{1, 2, \ldots, C\} \) for which they show the highest confidence score. Client \( i \) is assigned to cluster \( k \) such that:
   \[
   k = \arg\max_{1 \leq l \leq C} \text{conf}_{i,l},
   \]
The final outcome of this clustering method is at most $C$ clusters, where each cluster $k$ contains clients whose highest confidence score corresponds to class $k$. For each cluster $k$, the clients weights within the cluster are aggregated. We denote the model weights of client \( i \) as \( \theta_i \). The aggregated models weights for cluster $k$ are denoted as:
   \[
   \theta_k = \frac{1}{|K_k|} \sum_{i \in K_k} \theta_i,
   \]
where \( K_k \) is the set of clients in cluster \( k \), and \( |K_k| \) is the number of clients in that cluster. Finally, the updated model weights \( \theta_k \) are distributed back to each client \( i \) in cluster \( k \), which results in personalized models that adapt based on the cluster-specific aggregated knowledge.

\section{Proposed Method}\label{Section5}
\begin{figure}[h]
\centering
\includegraphics[width=1.05\textwidth]{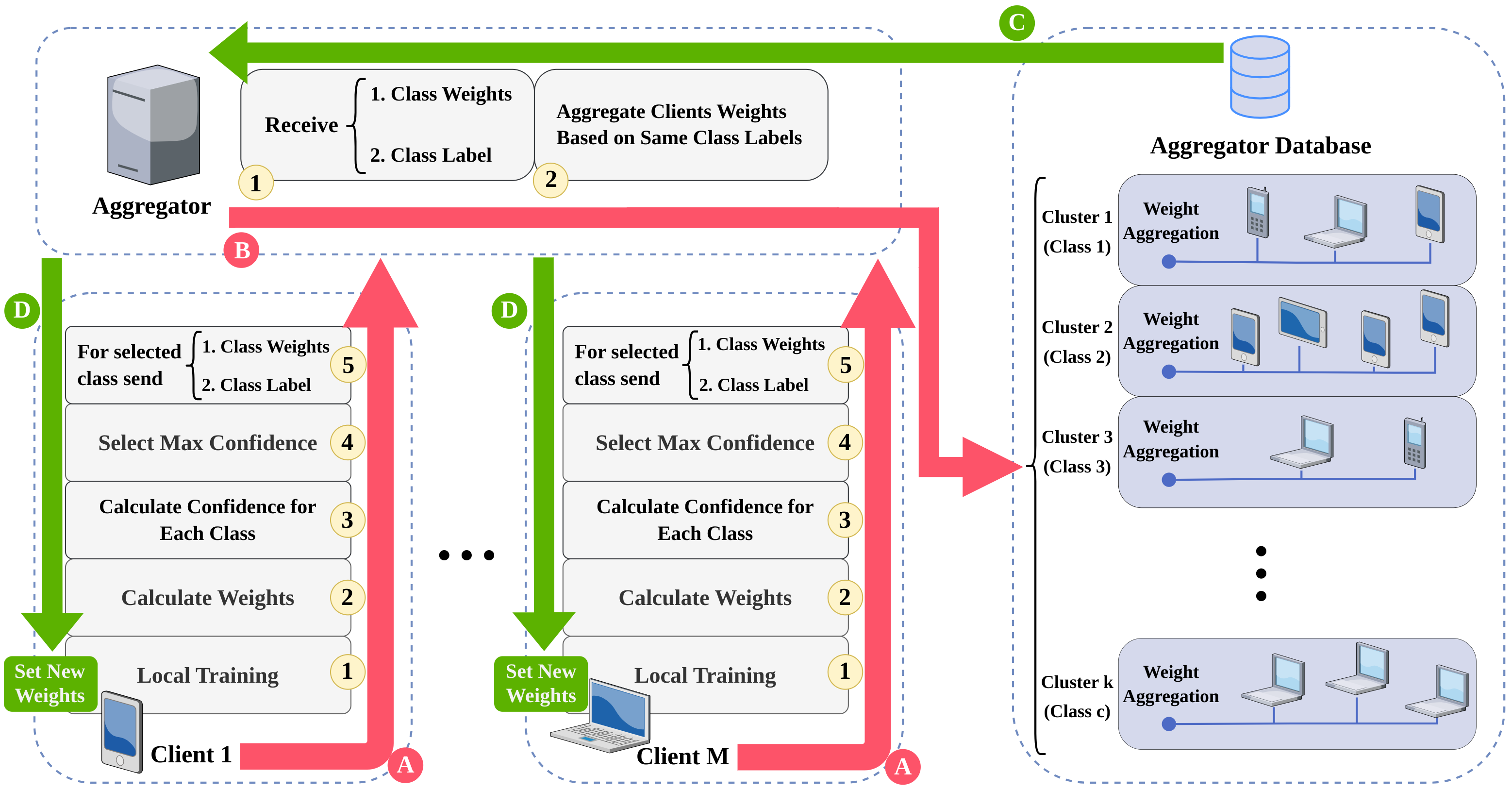}
\caption{Workflow of TPFL in one round of training.}\label{fig2}
\end{figure}

In order to address the heterogeneity of clients in FL settings, our proposed approach, TPFL, takes advantage of confidence-based clustering. Figure \ref{fig2} demonstrates the workflow of TPFL in a round of training. Each client in the TPFL goes through 4 phases in order to finish one round. Each phase may have a series of steps. Below, we will go over each phase individually.

\subsection{Phase A}
Phase A is where the process occurs on the local device. Phase A consists of 5 steps in total. In Phase A, in Step 1, each client starts training its local model using its private data. 

The client $m$ initiates the positive-polarity clauses \( C_j^+ \) and the negative-polarity clauses \( C_j^- \) as well as its weights \( W_m[c] \) for each class \( c \). The TM clauses are updated during each round using literals from the local training set \( D_{\text{train}} \). In Step 2 within each iteration, the client updates the clause weights for each class $c$ based on local data \( D_{\text{train}} \) and calculates a predicted output \( \hat{y} \). This process continues over several desired epochs to improve the classification accuracy of the local model. In Step 3, the confidence score is locally calculated using a separate set of data called \( D_{\text{conf}} \). This confidence stems from the voting machanism of TM algorithm. The more votes there are towards a class $c$, the more confident the model is in its prediction \( \hat{y} \). Afterwards, In Step 4 the local model selects the maximum confidence and sends a tuple to the aggregator, containing the weight vector \( W_m^c \) and its corresponding class $c$ to the aggregator. In this way each client sends only the weight vector \( W_m^c \) associated with the class $c$ for which it has the highest confidence. This selective transmission optimizes communication efficiency while ensuring that each client contributes meaningfully to the entire cluster in areas where it has the strongest expertise. Algorithm \ref{algo1} demonstrates Phase A in the local client.

\begin{algorithm}[H]
\caption{TPFL Client Algorithm}\label{algo1}
\textbf{Require:} Number of clients \( m \), number of classes \( C \), weight vector \( W_m[c] \), positive-polarity clauses \( C_j^+ \), negative-polarity clauses \( C_j^- \), local predicted class \( \hat{y}[m] \).

\textbf{Input:} Local train set \( D_{\text{train}} \), confidence set \( D_{\text{conf}} \).

\textbf{Output:} Updated weight vector \(W_m^{{c}_{\text{max}}}\) for class \(c_{\text{max}}\) with highest confidence.

\begin{algorithmic}[1]
\State Initialize \( C_j^+ \), \( C_j^- \), and \( W_m[c] \) for each class \( c \in \{1, 2, \ldots, C\} \).
\For{local training epoch}
    \State Update clause weights \( W_m[c] \) based on \( D_{\text{train}} \).
    \State Compute local prediction:
    \[
    \hat{y}[m] = \arg\max_{1 \leq c \leq C} \left( \sum_{j=1}^{n/2} W_j^{c, +}C_j^{c, +} - \sum_{j=1}^{n/2} W_j^{c, -}C_j^{c, -} \right).
    \]
\EndFor
\State Compute and find confidence score for each class \( c \) on \( D_{\text{conf}} \):
\[
c_{\text{max}} = \arg\max_{1 \leq c \leq C} \left( \sum_{x \in D_{\text{conf}}} \left( \sum_{j=1}^{n/2} C_j^{c,+}(x) - \sum_{j=1}^{n/2} C_j^{c,-}(x) \right) \right).
\]
\State Send weight vector \( W_m^{{c}_{\text{max}}} \) for class \( c_{\text{max}} \) to aggregator.
\end{algorithmic}
\end{algorithm}

\subsection{Phase B}
Phase B occurs on the aggregator, which only consists of retrieving the class $c$ that client $m$ has the highest confidence and its corresponding weight vector \( W_m[c] \). The aggregator's role in TPFL, unlike single-center FL, is to dedicate a single cluster $k$ to class $c$ such that the number of clusters is determined by the number of classes. 

\subsection{Phase C}
Initially, if a cluster has no previous weights, the weight vector retrieved from the first client is set as the initial weight for that cluster. As additional clients send their weights for the same class, the aggregator performs a summation of these weights, accumulating the updates from multiple clients. Once all clients have sent their weight updates to the aggregator, the aggregator computes the average weight for each cluster by dividing the accumulated weight by the number of clients that contributed to that cluster.
\subsection{Phase D}
Ultimately, the aggregator distributes the updated weight vectors \( W_{\text{cluster}}^k \) to the corresponding clients, providing each client with the new global weights. Each client is then evaluated on its test data to assess the performance improvements. Afterwards, the evaluation result of each model is received by the server. Algorithm \ref{algo2} shows Phase B, Phase C  and Phase D in the aggregator.

\begin{algorithm}[H]
\caption{TPFL Aggregator Algorithm}\label{algo2}
\textbf{Require:} Class index \( c \) corresponding to a cluster \( k \), weight vector \( W_m^{{c}_{\text{max}}} \) from client \( m \) for class \( c_{max} \), aggregated weight \( W_{\text{cluster}}^k \) for cluster corresponding to class \( c \), number of clients \( M_{\text{cluster}}^k \) in the cluster \( k \) for class \( c \).

\textbf{Input:} Received weight vectors \( W_m^{{c}_{\text{max}}} \) and its corresponding class \( c_{max} \) from client \( m \), local dataset \( D_{\text{test}} \).

\textbf{Output:} Aggregated weight vector \( W_{\text{cluster}}^k \) for each cluster \( k \).

\begin{algorithmic}[1]
\For{each round \( r \)}
    \For{each client \( m \)}
        \State \( k \leftarrow c_{max}\) \Comment{class \( c_{max} \) determines cluster \( k \) for \( W_m \)}
        \If{weights \( W_m^{{c}_{\text{max}}} \) are initial weights in cluster \( k \)}
            \State Initialize cluster weights \( W_{\text{cluster}}^k \) for cluster \( k \): 
            \[
            W_{\text{cluster}}^k \leftarrow W_m^{{c}_{\text{max}}}
            \]
        \Else
            \State Aggregate received weights:
            \[
            W_{\text{cluster}}^k \leftarrow W_{\text{cluster}}^k + W_m^{{c}_{\text{max}}}
            \]
        \EndIf
    \EndFor
    \For{each cluster \( k \)}
        \State Compute average of weights in the cluster:
        \[
        W_{\text{cluster}}^k \leftarrow \frac{1}{|M_{\text{cluster}}^k|} W_{\text{cluster}}^k
        \]
    \EndFor
    \State Send updated weight vectors \( W_{\text{cluster}}^k \) to corresponding clients of each cluster.
    \State Receive clients evaluation results on their \( D_{\text{test}} \).
\EndFor
\end{algorithmic}
\end{algorithm}

\subsection{TPFL Complexity}
As it can be seen in algorithm \ref{algo2}, the entire federation runs for 
\( R \) rounds, resulting in \( O(R)\) time complexity. Each round consists of two main iterations. The first part is the training of \( M \) clients, which has \( O(RM)\) complexity. The second part is the calculation of the average weights of \( K\) clusters, which has \(O(RK)\) complexity. Therefore, the TPFL time complexity can be shown as \( O(R (M+K))\).

\section{Experimental Results}\label{Section6}
In this section, we will go over the experimental results in detail. We will discuss experiment setups, datasets for the TPFL, the TPFL training settings and what baselines are used in order to make a comparison between the TPFL and other PFL techniques. In addition, we will also provide the evaluation metrics and the technique that we used to make our data non-IID. Our results are available in our Github repository.\footnote{\href{https://github.com/russelljeffrey/TPFL}{TPFL Github repository}}

\subsection{Experimental Setups}
\begin{figure}[h]
\centering
\includegraphics[width=1.0\textwidth]{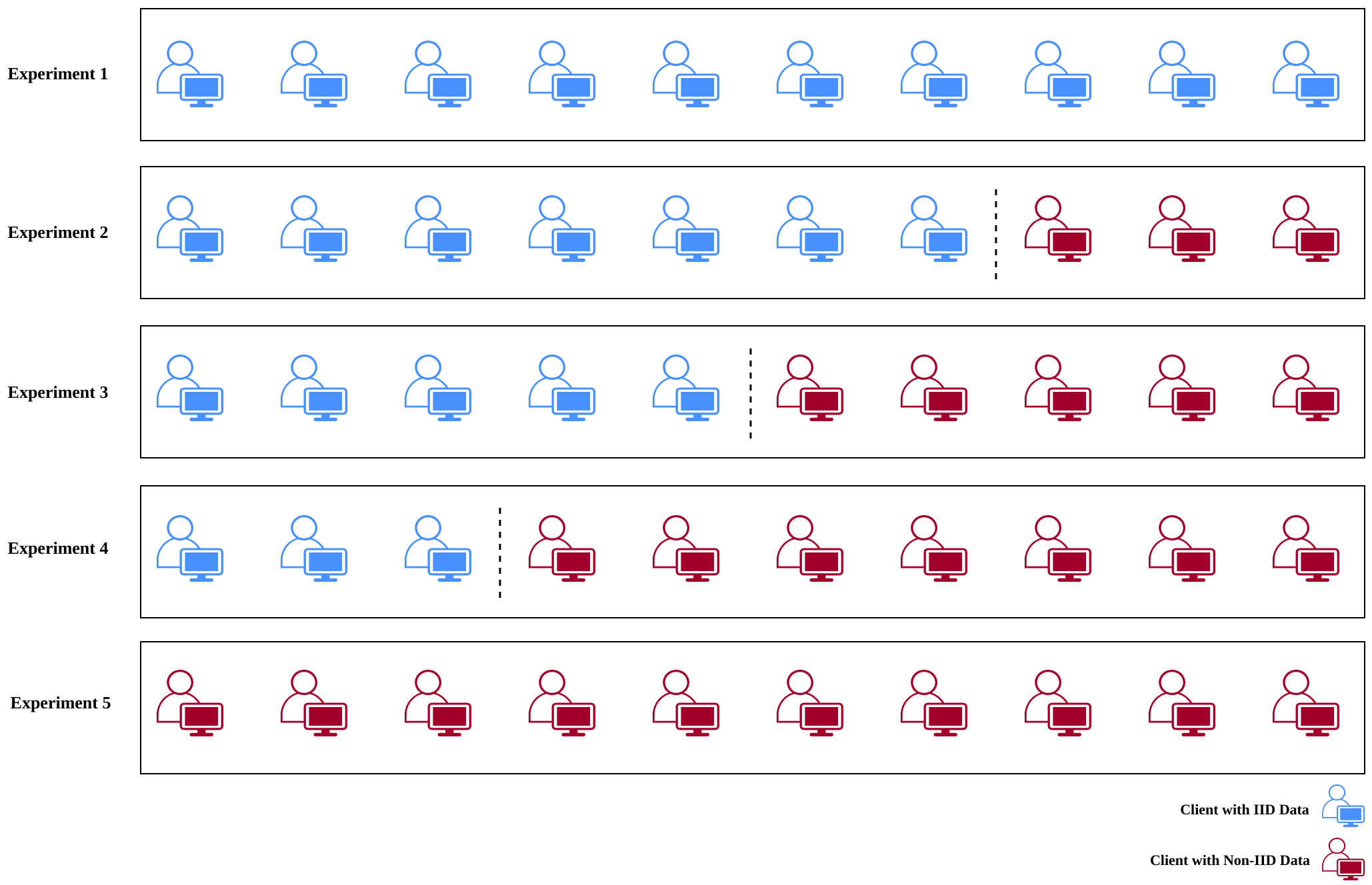}
\caption{Client data partitioning for TPFL under 5 different experimental setups}\label{fig3}
\end{figure}

To assess TPFL's performance under different data distribution scenarios, we conducted experiments with varying levels of IID and non-IID partitioning. Experiment 1 used fully IID partitioning, while experiments 2 to 5 progressively increased non-IID partitioning: 75\% IID and 25\% non-IID in experiment 2, 50\% IID and 50\% non-IID in experiment 3, and so on. Figure \ref{fig3} demonstrates the data partitioning for clients under 5 different experimental setups. Thus, for experiment 3, if we had 100 clients, 50 clients would use IID partitioning while the other half would use non-IID partitioning scheme.

\subsection{Datasets}
For training and evaluating the TPFL, we used three benchmarking datasets; MNIST, Fashion MNIST and Federated Extended MNIST (FEMNIST) datasets. However, it must be noted that in reality we used EMNIST instead of directly working with FEMNIST dataset. This is because, firstly, MNIST and Fashion MNIST were not available in frameworks that provide direct access to FEMNIST. Therefore, we created an independent function in the TPFL framework in order to create either IID or non-IID data based on class out of any dataset the user desires. Thus, enabling the user to just download any dataset they wish in order to create IID or non-IID dataset for clients in the federation (e.g. CIFAR10 and CIFAR100 are possible to be downloaded). Secondly, such frameworks are not constantly updated by maintainers so that it can be compatible with the latest version of Tensorflow (e.g. the LEAF framework \cite{w32} still requires Tensorflow 1). From here on in this paper, FEMNIST dataset is the partitioned data resulted from EMNIST. Table \ref{Dataset-Table} shows the details of the three datasets.

\begin{table}[!h] 
\caption{Summary of Datasets used in TPFL}
\label{Dataset-Table}
\begin{tabular}{|c|c|c|c|c|}
\hline 
\textbf{Dataset} & \textbf{Number of Images} & \textbf{Size} & \textbf{Classes} & \textbf{Format} \\ 
\hline
MNIST & 70,000 & 28 x 28 & 10 digits & Gray-scale \\
\hline
Fashion-MNIST & 70,000 & 28 x 28 & 10 items & Gray-scale \\
\hline
FEMNIST (by class) & 132,000 & 28 x 28 & 62 (letters, digits) & Gray-scale \\
\hline
\end{tabular}
\end{table}

\subsection{Data Partitioning} 
Both IID and non-IID data partition schemes are used in TPFL. Depending on different experimental setups, we used Dirichlet distribution to partition data into randomly shuffled IID and non-IID data \cite{w33}. Dirichlet distribution uses a parameter $\alpha$ in order to define the severity of distribution. We used $\alpha = 10000$ for IID data and $\alpha = 0.05$ for non-IID data. Each TPFL client takes advantage of three different sets from to the end of training rounds; a training set for training the local model, a test set for the evaluation of local model's performance and finally a confidence set so that the model's confidence can be evaluated on. 

\subsection{Evaluation Metrics}
The purpose of the proposed TPFL method is personalization based on clustering. Thus, the evaluation of the TPFL requires the overall performance of models in all clusters. We used classification accuracy as the metrics for the TPFL method. That is, the average accuracy of all models throughout all the existing clusters.
\subsection{Training Setups}
The TM algorithm relies on setting hyperparameters that are not similar to DL-based methods. We used 100 clients for all datasets under all 5 experiments. We trained each client on 10 local epochs and evaluated all clients for 10 rounds. We used 300 clauses for MNIST and 500 clauses for the remaining datasets and set sensitivity to 10 and feedback threshold T to 1000 during training. We also used 30000 randomly shuffled samples of each dataset for the train set, 15000 samples for the test set and similarly another 15000 samples for the confidence set. Table \ref{Conf-table} shows a summary of model configuration for each dataset.

\begin{table}[]
\caption{Summary of client configuration for TPFL.}
\label{Conf-table}
\begin{tabular}{|c|c|c|c|}
\hline
\textbf{Hyperparameter} & \textbf{MNIST} & \textbf{FashionMNIST} & \textbf{FEMNIST} \\ \hline
Train set & 30000 & 30000 & 30000 \\ \hline
Test set & 15000 & 15000 & 15000 \\ \hline
Confidence set & 15000 & 15000 & 15000 \\ \hline
Clause & 300 & 500 & 500 \\ \hline
Sensitivity & 10 & 10 & 10 \\ \hline
Feedback Threshold (T) & 1000 & 1000 & 1000 \\ \hline
Local epoch & 10 & 10 & 10 \\ \hline
Round & 10 & 10 & 10 \\ \hline
\end{tabular}
\end{table}

\subsection{Baselines}
We chose 5 baseline methods in order to compare the proposed TPFL with. Table \ref{baseline-summary} demonstrates the specification summary of these baselines. These methods consist of both vanilla FL and PFL. Moreover, the table also shows non-IID data generation has increasingly favored the Dirichlet distribution in recent years. The implementation details of these methods is as follows: 

\begin{enumerate}
\item \textbf{FedAvg}: This was the pioneering method that is still considered one of the most important baselines in the FL research community \cite{w34}. 
\item \textbf{FedProx}: Another important baseline proposed by Li et al. in \cite{w35}. We reimplemented this baseline using  proximal term 0.1.
\item \textbf{FLIS}: This approach uses inference similarity for clustering. They have provided numerous model architectures in their codebase. We used their codebase to reimplement their approach using only a simple CNN model. It is worth noting that their work is categorized into two approaches; FLIS Hierarchical Clustering (FLIS HC) and FLIS Dynamic Clustering (FLIS DC) \cite{w19}.
\item \textbf{IFCA}: We used their codebase to reimplement their work on only the FashionMNIST dataset since the other two datasets that we used were already implemented by the authors of IFCA \cite{w12}.
\item \textbf{FedTM}: This work is one of the most recent baselines for the TM algorithm in the FL setting without any personalization. No implementations were carried out for this work since the results were presented in FedTM \cite{w29}.
\end{enumerate}

We chose a range of popular works that are based on both vanilla FL and PFL. In addition, we included both DL-based works and a TM-based method in order to show a general overview of the TM algorithm potential in FL settings. Given the TM numerous advantages, its comparison with DL-based methods in FL context can shed a light on TM future works.

\begin{table}[]
\caption{Summary of baseline methods specifications.}\label{baseline-summary}
\begin{tabular}{|c|c|c|c|c|c|}
\hline
\textbf{Baseline} &
  \textbf{Year} &
  \textbf{PFL} &
  \textbf{Datasets} &
  \textbf{Reimplementation} &
  \textbf{Non-IID} \\ \hline
FedAvg \cite{w34} &
  2016 &
  No &
  CIFAR-10, MNIST &
  Full &
  \begin{tabular}[c]{@{}c@{}}LEAF\\ Library\end{tabular} \\ \hline
FedProx \cite{w35} &
  2018 &
  No &
  MNIST &
  Full &
  \begin{tabular}[c]{@{}c@{}}LEAF\\ Library\end{tabular} \\ \hline
FLIS DC \cite{w19} &
  2022 &
  Yes &
  \begin{tabular}[c]{@{}c@{}}CIFAR-100, CIFAR-10,\\ SVHN, FMNIST\end{tabular} &
  Partial &
  \begin{tabular}[c]{@{}c@{}}Dirichlet\\ Distribution\end{tabular} \\ \hline
FLIS HC \cite{w19} &
  2022 &
  Yes &
  \begin{tabular}[c]{@{}c@{}}CIFAR-100, CIFAR-10,\\ SVHN, FMNIST\end{tabular} &
  Partial &
  \begin{tabular}[c]{@{}c@{}}Dirichlet\\ Distribution\end{tabular} \\ \hline
IFCA \cite{w12} &
  2020 &
  Yes &
  \begin{tabular}[c]{@{}c@{}}MNIST, FEMNIST,\\ CIFAR-10\end{tabular} &
  Partial &
  NA \\ \hline
FedTM \cite{w29} &
  2023 &
  No &
  \begin{tabular}[c]{@{}c@{}}MNIST, FMNIST,\\ FEMNIST\end{tabular} &
  None &
  \begin{tabular}[c]{@{}c@{}}Dirichlet\\ Distribution\end{tabular} \\ \hline
\end{tabular}
\end{table}

\subsection{TPFL Results}
\begin{figure}[h!]
\centering
\begin{subfigure}[b]{0.3\textwidth}
\includegraphics[width=\textwidth]{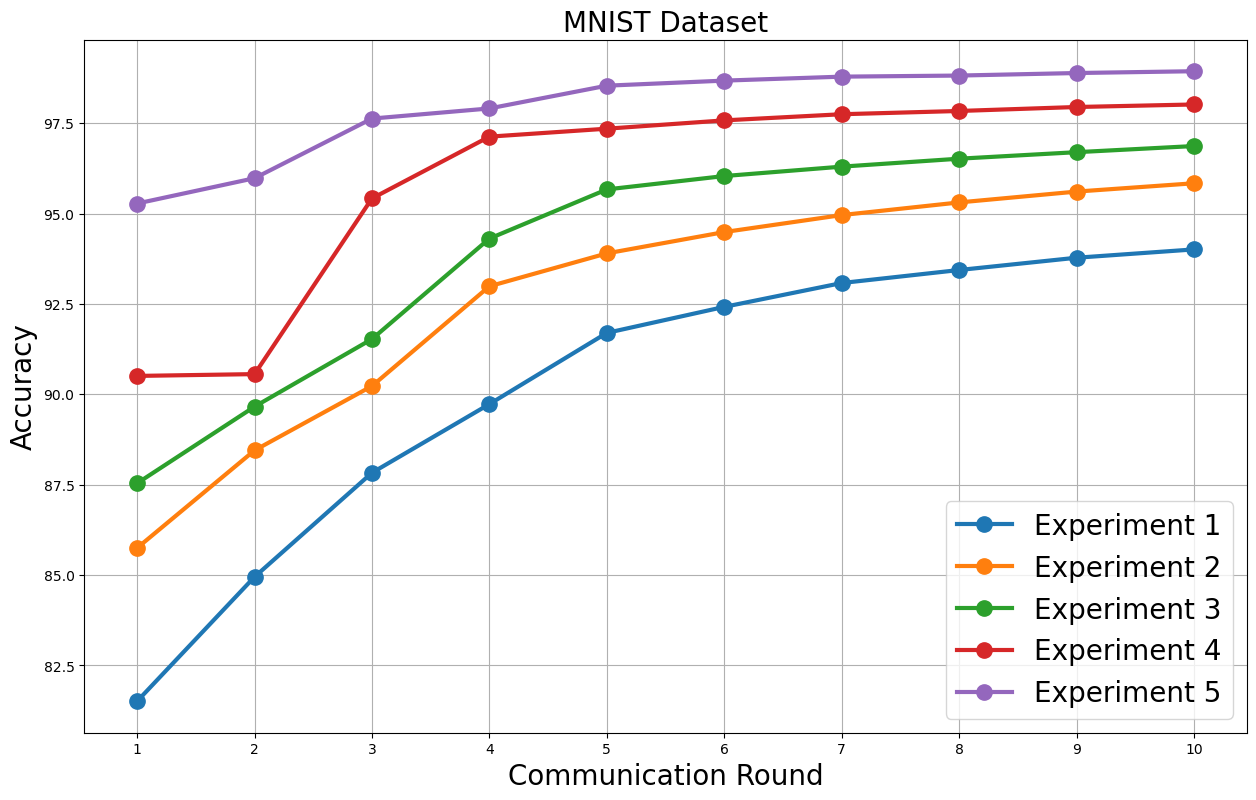}
\caption{MNIST convergence analysis under 5 experimental setups.}
\end{subfigure}
\hfill
\begin{subfigure}[b]{0.3\textwidth}
\includegraphics[width=\textwidth]{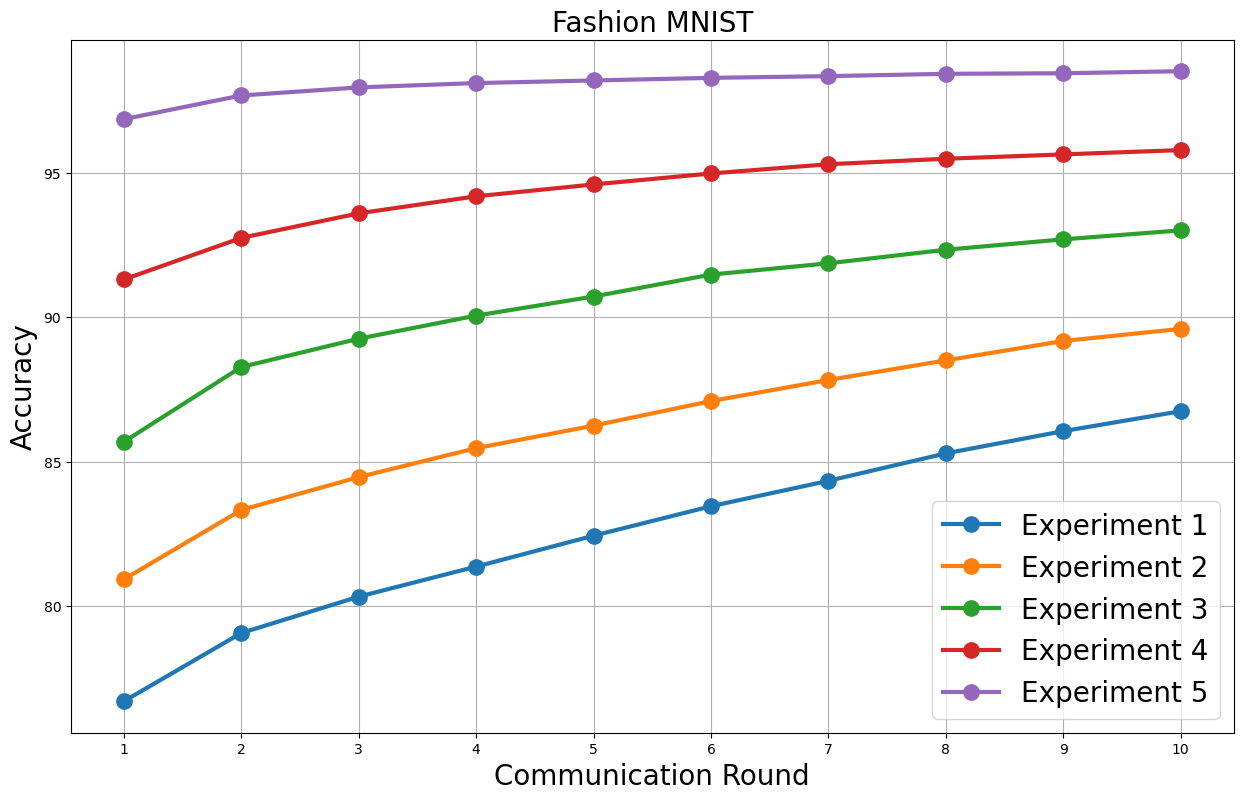}
\caption{FashionMNIST convergence analysis under 5 experimental setups.}
\end{subfigure}
\hfill
\begin{subfigure}[b]{0.3\textwidth}
\includegraphics[width=\textwidth]{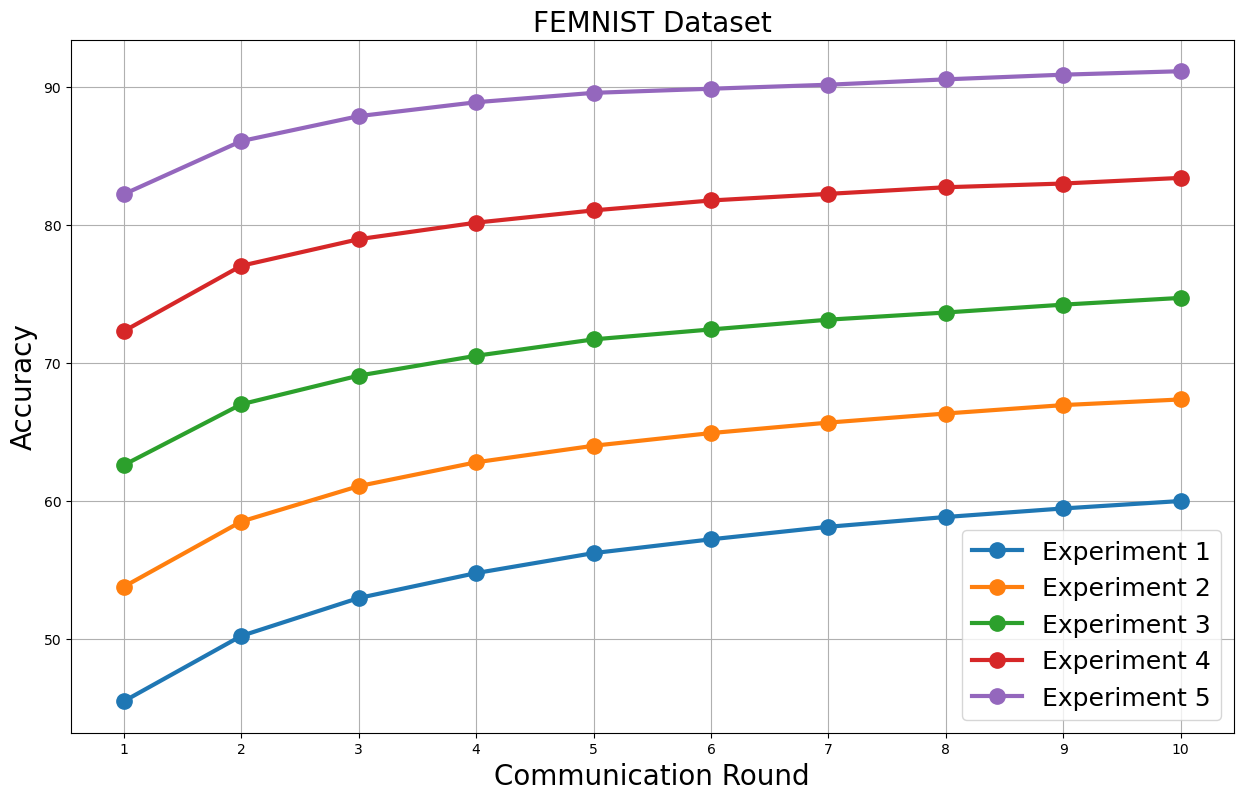}
\caption{FEMNIST convergence analysis under 5 experimental setups.}
\end{subfigure}
\caption{TPFL convergence analysis in terms of test accuracy under 5 experimental setups for MNIST, FashionMNIST, and FEMNIST.}\label{fig4}
\end{figure}

\begin{figure}[h!]
\centering
\begin{subfigure}[b]{0.3\textwidth}
\includegraphics[width=\textwidth]{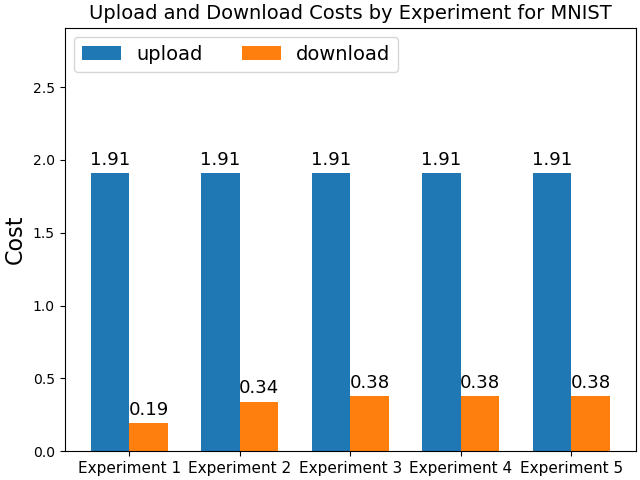}
\caption{MNIST communication costs under 5 experimental setups.}
\end{subfigure}
\hfill
\begin{subfigure}[b]{0.3\textwidth}
\includegraphics[width=\textwidth]{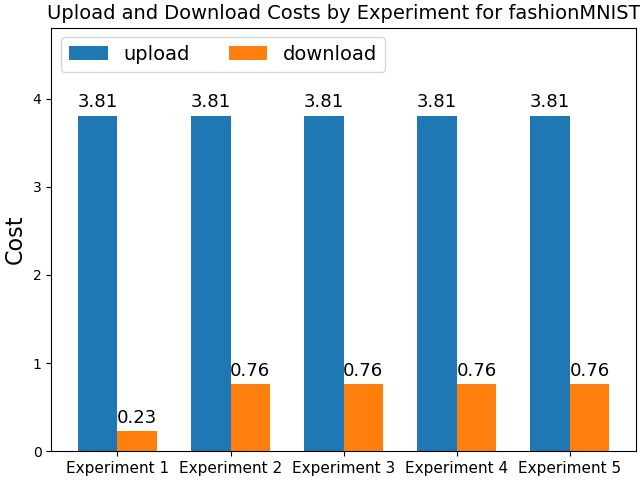}
\caption{FashionMNIST communication costs under 5 experimental setups.}
\end{subfigure}
\hfill
\begin{subfigure}[b]{0.3\textwidth}
\includegraphics[width=\textwidth]{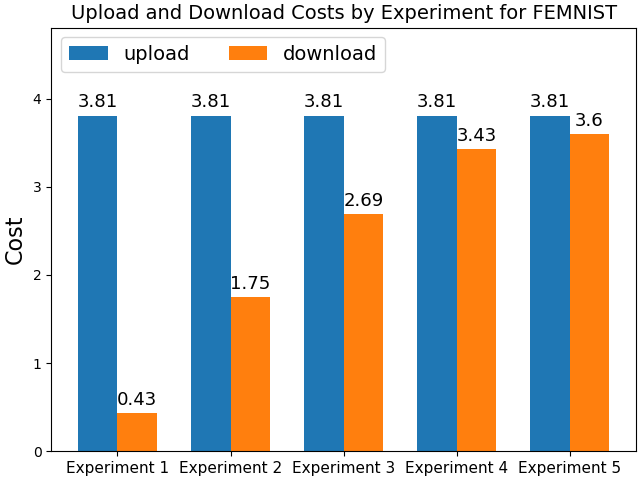}
\caption{FEMNIST communication costs under 5 experimental setups.}
\end{subfigure}
\caption{TPFL communication costs under 5 experimental setups for MNIST, FashionMNIST, and FEMNIST.}\label{fig5}
\end{figure}

\scalebox{0.65}{
\begin{talltblr}[
 caption={Test accuracies and communication costs of TPFL under 5 experimental setups.}\label{TPFL-table}
]{
 colspec={ccccccc},
 rowspec={Q[gray!20]QQ[gray!10]QQ[gray!10]QQ[gray!10]Q},
 vline{2-7} = {0-1}{0.3pt,gray!50},
 vline{2-7} = {2-7}{0.3pt,gray!30},
 hline{1,2,8} = {0.1pt,azure5}
}
\SetCell[r=2]{c}Experiment & \SetCell[c=2]{t}MNIST & & \SetCell[c=2]{c}Fashion-MNIST & &\SetCell[c=2]{c}FEMNIST \\
& {Accuracy\\(\%)} & {Upload/Download cost \\in 10 rounds(MB)} & {Accuracy\\(\%)} & {Upload/Download cost \\in 10 rounds(MB)} & {Accuracy\\(\%)} & {Upload/Download cost \\in 10 rounds(MB)} \\
Experiment 1  & 94.07 & 1.91/0.19 & 86.75 & 3.81/0.23 & 60.01 & 3.81/0.43 \\
Experiment 2  & 95.83 & 1.91/0.34 & 89.59 & 3.81/0.76 & 67.34 & 3.81/1.75 \\
Experiment 3  & 96.87 & 1.91/0.38 & 93.00 & 3.81/0.76 & 74.72 & 3.81/2.69 \\
Experiment 4  & 98.02 & 1.91/0.38 & 95.78 & 3.81/0.76 & 83.42 & 3.81/3.43 \\
Experiment 5  & \textbf{98.94} & 1.91/0.38 & \textbf{98.52} & 3.81/0.76 & \textbf{91.16} & 3.81/3.6 \\
\\
\end{talltblr}}
Table \ref{TPFL-table} shows the table test accuracy of the proposed TPFL method and the entire 10 rounds communication cost under 5 different experimental setups. As it can be seen, the highest accuracy under 5 different experimental setups belongs to experiment 5 when all models follow the non-IID partitioning scheme. Also, figure \ref{fig4}	 demonstrates that, regardless of the clients' dataset, the test accuracy in the starting round in experiment 5 is always greater than all other experimental setups.
\\
\\
\scalebox{0.69}{
\begin{talltblr}[
 caption={Summary of performance comparison between TPFL and baseline methods.}
 \label{baseline-table}
]{
 colspec={ccccccc},
 rowspec={Q[gray!20]QQ[gray!10]QQ[gray!10]QQ[gray!10]QQ[gray!10]Q},
 vline{2-7} = {0-1}{0.3pt,gray!50},
 vline{2-7} = {2-7}{0.3pt,gray!30},
 hline{1,2,10} = {0.1pt,azure5}
}
\SetCell[r=2]{c}Baselines & \SetCell[c=2]{t}MNIST & & \SetCell[c=2]{c}Fashion-MNIST & &\SetCell[c=2]{c}FEMNIST \\
& {Accuracy\\(\%)} & {Communication cost \\per model (MB)} & {Accuracy\\(\%)} & {Communication cost \\per model (MB)} & {Accuracy\\(\%)} & {Communication cost \\per model (MB)} \\
FedAvg \cite{w34}  & 73.34 & 1.64 & 61.46 & 1.64 & 50.48 & 1.74 \\
FedProx \cite{w35}  & 94.12 & 1.64 & 66.80 & 1.64 & 57.14 & 1.74 \\
FLIS DC \cite{w19}  & 97.65 & 0.41 & 79.22 & 0.41 & 52.80 & 0.43 \\
FLIS HC \cite{w19}  & 90.24 & 0.41 & 72.65 & 0.41 & 58.20 & 0.43 \\
IFCA \cite{w12}  & 95.25 & 24.78 & 85.62 & 24.78 & 87.89 & 49.98 \\
FedTM \cite{w29}  & 94.72 & 0.04 & 82.64 & 0.048 & 70.24 & 0.74 \\
TPFL & \textbf{98.94} & 0.019 & \textbf{98.52} & 0.038 & \textbf{91.16} & 0.038 \\
\\
\end{talltblr}}
This evident pattern clearly proves the relationship between the data distribution of all clients and their confidence. In other words, the more skewed the data distribution, the more confident the model towards that class, and the more confident the model, the faster their convergence time. Hence, the convergence time for our confidence-based clustering method is far less when the clients' partitioning scheme is severely non-IID. On the other hand, the models tend to be less confident when data distribution is uniformly IID, which is a clear indication that the entire federation requires much more time in order to become more confident and reach high-accuracy results. For such circumstances, increasing the local clients' training epochs or even increasing the total round number of federation can be beneficial so that high-accuracy results can be achieved. 
Figure \ref{fig5} shows the communication cost of all three datasets under 5 different experimental setups for a total of 10 rounds. The download cost is far lower than the upload cost, however, the more classes the dataset has, the more probable it is for the download cost to be more. Therefore, it can be seen for the FEMNIST dataset under experiment 5 that the download cost progressively becomes closer to the upload cost

\subsection{Comparison}
Clients in the FL settings are normally evaluated under non-IID data partitioning schemes. Therefore, We compared the baseline methods with the proposed TPFL method under non-IID data partitioning. Hence, our comparison is between the best results of each baseline with non-IID data partitioning and TPFL under experiment 5. Table \ref{baseline-table} shows the summary of our comparison. As it can be seen, the TPFL model outperforms all baselines in all datasets followed by IFCA and then FedTM. It is worth mentioning that TFPL implementation was carried out only in 10 rounds, the least number of rounds among all baseline methods. This phenomenon illustrates the positive effect of cluster-based personalization within TPFL setting. \\
This evident result shows the difference between DL-based models and the TM-based models. Based on the obtained results and their comparison with baseline models in figure \ref{fig4}, it can be understood that the TPFL approach can be utilized in a one-shot learning fashion, in which each experimental setup is carried out only in one round. However, it is advisable to increase the number of local training epochs in order to make up for absence of multiple rounds of training. \\
As it can be seen from table \ref{baseline-table}, in case of high-class datasets, such as FEMNIST, TPFL has a key advantage in upload communication efficiency per model. This phenomenon stems from the fact that DL-based techniques necessitate the transmission of a full weight vector for the entire model, while in TPFL only the weight vector corresponding to the class with the highest confidence score is uploaded from each local client. As shown in figure \ref{fig5}, in case of download communication cost, TPFL is still more efficient when it comes to communication cost. However, assuming a federation of 100 clients with CIFA100 dataset, the worst case scenario for the TPFL is that the download cost would be equal to the upload cost as no client in the federation shares a similar class to aggregate their weights. Figure \ref{fig5} shows this correlation as the download communication cost progressively goes higher when clients have to deal with more classes of data.

\section{Conclusion}
With the advent of FL algorithms, numerous works have been put forward to maintain the users privacy. However, vanilla FL suffers from numerous disadvantages. Clustering-based PFL is considered a promising solution to address the shortcomings of the vanilla FL. Given the beneficial and explainable voting mechanism it has, the TM can be a great candidate for clustering-based PFL. In this work, we propose a personalization method called Tsetlin-Personalized Federated Learning (TPFL) that leverages the voting mechanism of the TM in order to cluster clients based on their confidence towards a class. The results of TPFL compared with other FL and PFL baselines demonstrate that TPFL outperforms all baselines for MNIST, FashionMNIST and FEMNIST datasets. A promising future work that can be built upon TPFL is the utilization of innovative clustering strategies such as thresholding strategies for defining a confidence threshold and multiclass selection of class weights | sharing weights of more than one class | so that a client can be assigned to multiple clusters.

\section{Declarations}
\subsection{Competing Interests}

The authors have no relevant financial or non-financial interests to disclose.

\subsection{Authors Contribution Statement}
Author Rasoul Jafari Gohari conceptualized and implemented the proposed method while all authors collaboratively structured the the article, prepared the manuscript, analyzed data, reviewed and edited drafts, and finally approved the publication.

\subsection{Code availability}
All the codes and implementations are available via our \href{https://github.com/russelljeffrey/TPFL}{Github repository}.

\subsection{Funding}
The authors declare that no funds, grants, or other support were received during the preparation of this manuscript.
\subsection{Compliance with Ethical Standards}
All authors have followed the ethical standards in conducting and reporting this research, as outlined by the Springer Nature Code of Conduct for Authors.

\end{document}